\documentclass[aps,prd,preprint,tightenlines,superscriptaddress,
amsfonts,amssymb,amsmath,byrevtex,showpacs,nofootinbib]{revtex4-2}
\usepackage[polish,english]{babel}
\usepackage[utf8]{inputenc}
\usepackage[T1]{fontenc}
\usepackage{ulem}
\usepackage{latexsym}
\usepackage{graphicx}
\usepackage{geometry}
\usepackage{epstopdf}
\usepackage{subfig}
\usepackage{color}

\usepackage{xcolor}
\usepackage{pdfsync}
\usepackage{fancyhdr}
\usepackage{makeidx}
\usepackage[breaklinks,colorlinks,backref]{hyperref}
\hypersetup{
    colorlinks, %set true if you want colored links
    linktoc=all, %set to all if you want both sections and subsections linked
    linkcolor=red,  %choose some color if you want links to stand out
  citecolor=hyptxt,
  urlcolor=blue}
  \hypersetup{
  citebordercolor=,
  filebordercolor=green,
  linkbordercolor=blue
}
\definecolor{hyptxt}{rgb}{0.7, 0.4, 0.9}
\usepackage{bibtopic}
\newcommand{\bea}{\begin{eqnarray}}
\newcommand{\eea}{\end{eqnarray}}

\newcommand{\dR}{\mathbb R}

\newcommand{\be}{\begin{equation}}
\newcommand{\ee}{\end{equation}}
\newcommand{\CR}{{\mathcal{R}}}

\newcommand{\R}{\mathbb R}

\newcommand{\ket}[1]{|\kern.3ex#1\kern.3ex\rangle}
\newcommand{\bra}[1]{\langle\kern.3ex #1 \kern.3ex|}
\newcommand{\scalar}[2]{\langle\kern.3ex #1 \kern.3ex|\kern.3ex#2\kern.3ex\rangle}
\newcommand{\norm}[1]{\|\kern.3ex#1\kern.3ex \|}

\newcommand{\RNumb}{\mathbb{R}}

\newcommand{\UnitOp}{\hat{1\kern-4.75pt 1}} % matrix unit
\newcommand{\MatUnit}{1\kern-3pt 1} % matrix unit
\newcommand{\Group}[1]{\textrm{#1}} % group assignment
\newcommand{\Aff}{\textrm{Aff}(\RNumb)}
\newcommand{\Bra}[1]{\langle #1 \vert} % bra
\newcommand{\Ket}[1]{\vert #1 \rangle} % ket
\newcommand{\BraKet}[2]{\langle #1 \vert #2 \rangle} % bra(c)ket
\newcommand{\ScProd}[2]{( #1 \vert #2 )} % Scalar product

\newcommand{\Var}[1]{\mathrm{var}(#1)} %  symbol variance

\newcommand{\EOp}{\mathsf{E}\kern-1pt\llap{$\vert$}}   % PEv E operator
\newcommand{\WOp}{\hat{\mathsf{W}\kern-1pt\llap{$-$}}} % PEv W generator
\newcommand{\FOp}{\mathsf{F}\kern-1pt\llap{$\vert$}}   % QEv F operator
\newcommand{\StateSpace}[1]{\mathcal #1} % state space
\newcommand{\rmd}{\mathrm{d}} % font roman d
\begin{document}

\title{Quantum system ascribed to the Oppenheimer-Snyder \\ model of massive star}

\author{Andrzej G\'{o}\'{z}d\'{z}}
\email{andrzej.gozdz@umcs.lublin.pl}
\affiliation{Institute of Physics, Maria Curie-Sk{\l}odowska
University, pl.  Marii Curie-Sk{\l}odowskiej 1, 20-031 Lublin, Poland}

\author{Jan J. Ostrowski}
\email{Jan.Jakub.Ostrowski@ncbj.gov.pl}
\affiliation{Department of Fundamental Research, National Centre for Nuclear
  Research, Pasteura 7, 02-093 Warszawa, Poland}

\author{Aleksandra P\c{e}drak} \email{aleksandra.pedrak@ncbj.gov.pl}
\affiliation{Department of Fundamental Research, National Centre for Nuclear
  Research, Pasteura 7, 02-093 Warszawa, Poland}

\author{W{\l}odzimierz Piechocki} \email{wlodzimierz.piechocki@ncbj.gov.pl}
\affiliation{Department of Fundamental Research, National Centre for Nuclear
  Research, Pasteura 7, 02-093 Warszawa, Poland}

%%%%%%%%%%%%%%%%%%%%%%%%%%%%%%%%%%%%%%%%%%%%%%%%%%%%%%%%%%%%%%%%%%%%%%%%
\date{\today}
%%%%%%%%%%%%%%%%%%%%%%%%%%%%%%%%%%%%%%%%%%%%%%%%%%%%%%%%%%%%%%%%%%%%%%%%

\begin{abstract}
We quantize the Oppenheimer-Snyder model of black hole using the integral quantization method.
We treat spatial and temporal coordinates on the same footing both at classical and quantum levels.
Our quantization resolves or smears the singularities of the classical curvature invariants.
Quantum trajectories with bounces can replace singular classical ones. The considered quantum
black hole may have finite bouncing time. As a byproduct, we obtain the resolution of the gravitational
singularity of the Schwarzschild black hole at quantum level.
\end{abstract}

%\pacs{XXX}

\maketitle

\tableofcontents

%%%%%%%%%%%%%%%%%%%%%%
\section{Introduction}
%%%%%%%%%%%%%%%%%%%%%%

Cosmological models can be used to describe black holes after imposing
the condition that involves dealing with isolated objects. In the case of
spherically symmetric objects, this issue may be reduced to the problem of matching
the Schwarzschild spacetime with finite region of specific spacetime
\cite{WI1,WI2}. This idea was recently used to obtain the Oppenheimer-Snyder
 (OS) and Lema\^{i}tre-Tolman-Bondi (LTB) models of
isolated objects within one formalism (see \cite{Kwidzinski:2020xyd} and
references therein).  The former model concerns a spherical cloud of homogeneous
dust (pressureless matter), whereas the latter one deals with a spherical but
inhomogeneous cloud of dust.
However,  the merging condition leads sometimes to
complicated equations defining variables, which cannot be resolved analytically
but only numerically, creating additional difficulties in analyses (e.g., see
\cite{Kwidzinski:2020xyd}).

Different strategy of obtaining a description  of an isolated astrophysical object was
proposed recently  for LTB model \cite{LL1,LL2,LLB,Ed1}.  A metric of both interior and
exterior regions of isolated body is expressed in one coordinate system,  which
avoids the need for imposing the matching conditions across the interface
between matter and vacuum regions provided that certain functions are continuous across the boundary.
We apply this approach in the present paper.

In this article we present the quantum system ascribed to the OS model of collapsing
pressureless dust star.
For this purpose we use the so-called integral quantization (IQ) method  applied
quite recently to the quantization of the Schwarzschild spacetime \cite{Ola} and a
thin matter shell in vacuum \cite{Marcin}.

In this article, we disregard the black hole evaporation by the Hawking radiation.
Thus, the global mass of considered star is conserved during its evolution.

Following the idea presented in \cite{Ola}, we quantize not only spatial but also temporal coordinates.
Rationale for such an approach is the covariance of general relativity with respect to the transformations
of these  coordinates. Treating temporal and spatial coordinates on the same footing at quantum
level has enabled the construction of a consistent quantum theory.

One of the main issues addressed in the quantization of black holes is the singularity avoidance.
It was widely discussed in the literature. In the case of the OS black hole, various approaches
motivated by the loop quantum gravity have led to the resolution of the singularity of the classical model
(see, e.g., \cite{Jar,Jerzy,Jin,Kri,Mux,Tomasz,Pullin} and references therein).  There exist many
investigations done in the context of quantum geometrodynamics. See, for instance, \cite{Hajicek} and
references therein. On the other hand, there are few results obtained within the
IQ method based on coherent states \cite{Wlo,Cla}. The latter method is applicable even in the cases 
the canonical quantization has methodological problems \cite{Cla}. We extend this discussion in the conclusion 
section.

The paper is organized as follows: In Sec. II we recall the formalism of
spherically symmetric spacetime specialized to black holes. The singularities
and horizon issues are exhibited. The case of the dust black hole is made
specific.  Sec. III is devoted to the quantization of the OS model. It includes
recalling an essence of the IQ quantization and presents the quantum dynamics of
considered black hole. In particular, we discuss the fate of the classical
singularities at quantum level. We reduce the results to the case of the quantum
Schwarzschild black hole in Sec. IV. We conclude in Sec. V.
App. A recalls general solution for the dynamics of the LTB spacetime. App. B specifies
the transformation between two coordinate systems applied  in our paper. The two state spaces used
in the affine quantization are discussed  in App. C.

\noindent In the following we choose $\;G = c =1 = \hbar\;$ except where otherwise noted.

%%%%%%%%%%%%%%%%%%%%%%%%%%%%%%%%%%%%%%%%%
\section{Spherically symmetric spacetime}
%%%%%%%%%%%%%%%%%%%%%%%%%%%%%%%%%%%%%%%%%

%%%%%%%%%%%%%%%%%%%%%%%%%%%%%%%%%%%
\subsection{The perfect fluid case}
%%%%%%%%%%%%%%%%%%%%%%%%%%%%%%%%%%%

We begin our considerations with recalling the metric and field equations of the
LTB model of black hole presented in the article \cite{LL1}, and describing a
general spherically symmetric perfect fluid.

The metric of the collapsing massive star of a spherically symmetric spacetime
in the coordinates $(t,r,\theta,\phi)$ (see, e.g., \cite{Wald}) and considered in
\cite{LL1} reads
\begin{equation}\label{eq1}
  \rmd s^2 = - \alpha^2 \rmd t^2 + \frac{(\alpha \sqrt{2 M/r + E}\,
 \rmd t + \rmd r)^2}{1+E} + r^2 \rmd \Omega^2 \, ,
\end{equation}
where $(t,r) \in \R \times \R_+$, with $ \R_+ := \{r \in \R~|~r > 0\}$, and where
$\alpha (t,r) > 0$ is the lapse function (specific to the $3+ 1$ decomposition
of the spacetime metric), whereas $\rmd \Omega^2 := \rmd \theta^2 +\sin^2\theta
\rmd \phi^2$. The so-called energy function $E(t,r)$, arbitrary up to the
constraint $E > -1$, is a measure of the energy of a shell at a radius $r$. The
so-called mass function $M(t,r)$, at some radius $r$, is defined to
be
\begin{equation}\label{eq6}
  M(t,r) = 4\pi\int_0^r \rho(t,\sigma) \sigma^2 \; d \sigma \, .
\end{equation}

The field equations have the form \cite{LL1}
\begin{equation}\label{eq2}
  \frac{\partial M}{\partial t} - \alpha\,\sqrt{\frac{2M}{r}+E} \;\;
\Big(\frac{\partial M}{\partial r} +  4\pi P r^2 \Big) = 0 \, ,
\end{equation}
\begin{equation}\label{eq3}
\frac{\partial E}{\partial t} - \alpha\,\sqrt{\frac{2M}{r}+E} \;\;
\Big(\frac{\partial E}{\partial r} + 2\, \frac{1+E}{\rho +P}\;
\frac{\partial P}{\partial r}   \Big) = 0 \, ,
\end{equation}
\begin{equation}\label{eq4}
  \frac{1}{\alpha}\;\frac{\partial \alpha}{\partial r}
+ \frac{1}{\rho + P}\; \frac{\partial P}{\partial r} = 0 \, ,
\end{equation}
\begin{equation}\label{eq5}
  f(\rho, P) = 0 \, .
\end{equation}
The energy-momentum tensor for a perfect fluid reads
\begin{equation}\label{eq7}
  T^{\mu\nu} = (\rho + P)\,n^\mu n^\nu + P g^{\mu\nu}\, ,
\end{equation}
where $P$ is the pressure, $\rho \neq -P$ is the energy density, $n^\mu$ is the
vector field tangent to the fluid, and where \eqref{eq5} is an
equation of state.

In the OS collapse model of black hole, we have two regions: a ``star region'' defined
by non-vanishing $\rho$ and a vacuum region with vanishing energy density. In this configuration,
the system is asymptotically flat and $M$ presents the global mass/energy of the system.
This energy is conserved during the evolution as a static spherically symmetric black hole
does not radiate gravitational waves.

Eqs. \!\eqref{eq2}--\eqref{eq5} require the specification of initial and
boundary conditions to be well defined. Additionally, one should impose the
boundary conditions at the interface between the matter that fills the interior
of the ball of matter and the vacuum exterior. For these important issues we
recommend Sec. \!III of Ref. \!\cite{LL1}.

The metric \eqref{eq1} is expressed in terms of a generalization of the
Painlev\'{e}-Gullstrand (PG) coordinates \cite{PP,AG}. In this choice of
coordinates, the radial coordinate $r$ is always spacelike.  We call the above
formalism, the spherically symmetric spacetime model in the PG coordinates.

%%%%%%%%%%%%%%%%%%%%%%%%%%
\subsection{The dust case}
%%%%%%%%%%%%%%%%%%%%%%%%%%

In what follows, we consider the dust case. It can be introduced by the
conditions
\begin{equation}\label{d1}
P=0\;\;,\;\;\alpha = 1\;,
  \end{equation}
  implemented into Eqs. \!\eqref{eq1}--\eqref{eq4}. As the result, the metric
  and field equations become
\begin{eqnarray}
 \label{d2}  \rmd s^2 &=& - \rmd t^2 + \frac{(\sqrt{2 M/r + E}\, \rmd t
+ \rmd r)^2}{1+E} + r^2 \rmd \Omega^2 \, , \\
\label{dd3} \frac{\partial M}{\partial t} &-& \sqrt{\frac{2M}{r}+E} \;\;
\frac{\partial M}{\partial r}  = 0  \, , \\
\label{dd4} \frac{\partial E}{\partial t} &-& \sqrt{\frac{2M}{r}+E} \;\;
\frac{\partial E}{\partial r} = 0 \, ,
\end{eqnarray}
where $M$ is defined by Eq. \!\eqref{eq6}, and where $t$ is the time measured by
an in-falling geodesic observer.

%%%%%%%%%%%%%%

The system \eqref{d2}--\eqref{dd4} can describe the interior and exterior regions of the LTB
dust star in the single PG coordinate patch.
Given the initial density profile with the density vanishing at some finite radius, the latter
region can be shown to be diffeomorphic to the Schwarzschild spacetime \cite{LLB}.
The so-called marginally bound model \cite{CLT}, considered in this article, is the case with $E =
0$ (see, App. A). The classical dynamics of this model is defined, due to \eqref{dd3}, by the
equation
\begin{equation}\label{dd5}
  \frac{\partial M}{\partial t} - \sqrt{\frac{2M}{r}} \;\;
\frac{\partial M}{\partial r}  = 0 \, .
\end{equation}
 This equation has a general implicit solution (as presented in \cite{LL1}):
\begin{equation}\label{DD5}
M = \mathcal{F}\left(\frac{2}{3}r^{3/2}+t\sqrt{M}\right) \;,
\end{equation}
where $\mathcal{F}$ is a function of integration coming from solving the equation with the method of characteristics.
In what follows we present a simpler but explicit, separable analytical solution to Eq. \!\eqref{dd5},
which is a particular case satisfying Eq. \!\eqref{DD5} (see, App. A for a general solution  in $(t,R)$ coordinates).

%%%%%%%%%%%%%%%%%

Making the assumption $M(t,r)=M_1(t)M_2(r)$, splits \eqref{dd5} into the two
equations:
\begin{eqnarray}
&& \label{eq:3d2}
\frac{\partial M_1}{\partial t} - \lambda M_1^{\frac{3}{2}}  =0 \, ,\\
&& \label{eq:4d2}
\frac{\partial M_2}{\partial r} - \lambda  \sqrt{\frac{M_2 r}{2}} =0 \, .
\end{eqnarray}
Direct integration of the equation \eqref{eq:3d2} gives the  solution
\begin{equation}
\label{eq:3a-d2}
\sqrt{M_1}= \frac{-2}{\lambda t + C_1} \ ,
\end{equation}
where $\lambda$ is the separation parameter and $C_1$ an arbitrary constant.

Similarly, integration of the equation \eqref{eq:4d2} gives the following
solution
\begin{equation}
\label{eq:4a-d2}
\sqrt{M_2}= \frac{\lambda}{3\sqrt{2}} r^{\frac{3}{2}} +C_2 \ ,
\end{equation}
where $C_2$ is an arbitrary constant.

Combining both factors $M_1$ and $M_2$, the full solution of the dust equation
can be written as
\begin{equation}
\label{eq:34d2}
\sqrt{M(t,r)}=\sqrt{M_1(t)}\sqrt{M_2(r)}
= \left(\frac{-2}{\lambda t + C_1}\right)
\left( \frac{\lambda}{3\sqrt{2}} r^{\frac{3}{2}} +C_2 \right) \, ,
\end{equation}
so that
\begin{equation}
\label{eq:34A-d2}
M(t,r) = \left(\sqrt{M_1(t)}\sqrt{M_2(r)} \right)^2
= \left\{ \left(\frac{-2}{\lambda t + C_1}\right)
\left( \frac{\lambda}{3\sqrt{2}} r^{\frac{3}{2}} +C_2 \right) \right\}^2 \, .
\end{equation}

Formally,  the derivation of the solution \eqref{eq:34A-d2} requires the restrictions of the domains
of the functions $\sqrt{M_1(t,r)}$ and $\sqrt{M_2(t,r)}$, which also implies the restriction for the domain
of $M(t,r)$. However, since the function \eqref{eq:34A-d2} fulfills  the condition of mass positivity for
arbitrary $(t,r)$, we take the mass \eqref{eq:34A-d2} to be defined on the full domain $(t,r)\in\RNumb\times \RNumb_+$,
except the singularities at $t_c = -C_1/\lambda$. This result can be mathematically  well defined by taking the
definition of the complex root in  the solutions to Eqs. \!\eqref{eq:3d2} and  \eqref{eq:4d2}.

%%%%%%%%%%%%%%%%%%%%%%%%%%
\subsection{Singularities}
%%%%%%%%%%%%%%%%%%%%%%%%%%

In the synchronous comoving coordinates $(\tau, R, \theta, \phi )$, the line element and field equations read (see,
e.g., \cite{PlebanskiKrasinski})

\begin{equation}\label{line2}
\rmd s^2 = -\rmd \tau^2 + \frac{\left(\frac{\partial r}{\partial R}\right)^2}{1+2E}\rmd R^2 + r^2 \rmd \Omega^2\;,
\end{equation}
\begin{eqnarray}
\label{com1} \left(\frac{\partial r}{\partial \tau}\right)^2 &=&\frac{2M}{r}+E \;,\\
\label{com2} 4\pi \rho &=& \frac{\partial M}{\partial R}\left(r^2\frac{\partial r}{\partial R}\right)^{-1}\;.
\end{eqnarray}
Eq. \!\eqref{com1}  describes the dynamics of the shells. Taking a square
root of this equation requires choosing a sign, which represents expanding or
collapsing shells for negative or positive sign, respectively.  The second
equation \eqref{com2}  defines the density, which can become singular under two conditions
leading to two different types of singularities. The $\partial r /\partial R = 0$ is usually
referred to as a shell crossing  and may be viewed as a weak singularity, inherent to the lack
of pressure in the matter model. The second one however, i.e. $r=0$, is a strong singularity and cannot
be avoided for the collapsing scenarios.
For the relationship between the line elements \eqref{d2} and \eqref{line2} expressed in two
different coordinate systems, see App B.

%%%%%%%%%%%%%%%%%%%%%%%%%%%%%%%%%%%%%%%%%%
\subsubsection{Shell crossing singularity}
%%%%%%%%%%%%%%%%%%%%%%%%%%%%%%%%%%%%%%%%%%
Let us find an explicit form of the condition  $\partial r /\partial R = 0$ for
the separable solution \eqref{eq:34A-d2}. We rewrite it as
\begin{equation}\label{Mrt}
M(t,r)  = \left(A r^{\frac{3}{2}}+B\right)^2 \;,
\end{equation}
where
\begin{equation}
A = \frac{-2\lambda}{\left(\lambda t+C_1\right)3\sqrt{2}} \;\;,\;\;B = \frac{-2C_2}{\lambda t+C_1}\;.
\end{equation}

Let us take some instance of time $t=0$ so that
\begin{equation}
M(t_i, r_i) = \left(\left(\frac{-2}{C_1}\right)\left(\frac{\lambda}{3\sqrt{2}}r_i^{\frac{3}{2}}+C_2\right)\right)^2
= \left(A_i r_i^{\frac{3}{2}}+B_i\right)^2 \;.
\end{equation}
We can use the freedom to choose our radial coordinate and set $R = r_i$, where $r_i$ is the areal radius at $t=0$.
We then have
\begin{equation}
M(0,R) = \left(A_i R^{\frac{3}{2}}+B_i\right)^2 \;.
\end{equation}
We also get from \eqref{Mrt}
\begin{equation}
r = \left(\frac{\sqrt{M}-B}{A}\right)^{\frac{2}{3}} \;.
  \end{equation}
Differentiating the above equations leads to:
\begin{equation}\label{par1}
\frac{\partial r}{\partial R} = \frac{1}{3A}r^{-\frac{1}{2}}M^{-\frac{1}{2}}\frac{\partial M}{\partial R}\;,
\end{equation}
and
\begin{equation}\label{par2}
\frac{\partial M}{\partial R} = 3\sqrt{M}A_iR^{\frac{1}{2}}\;.
\end{equation}

From the two equations above it is evident that the shell-crossing cannot occur as for our setup
we have

\begin{equation}
\frac{\partial r}{\partial R} = \frac{\lambda t +C_1}{3\sqrt{2}\, C_1} \neq 0~~~~\mbox{for}~~~~t \neq  t_c = - C_1/\lambda \;.
  \end{equation}
%

%%%%%%%%%%%%%%%%%%%%%%%%%%%%%%%%%%%%%%%%%%%
\subsubsection{Gravitational singularities}
%%%%%%%%%%%%%%%%%%%%%%%%%%%%%%%%%%%%%%%%%%%
The existence of gravitational singularity is usually signalized by the blow-up of the curvature scalars.
The Kretschmann scalar can be decomposed in the following way:
\begin{equation}\label{KKK}
K = R_{\mu \nu \lambda \sigma}R^{\mu \nu \lambda \sigma}
= C_{\mu \nu \lambda \sigma}C^{\mu \nu \lambda \sigma}
+ 2R_{\mu \nu}R^{\mu \nu} - \frac{1}{3}\CR^2 \;,
\end{equation}
where $R_{\mu \nu \lambda \sigma}$, $C_{\mu \nu \lambda \sigma}$, $R_{\mu \nu}$
and $\CR$ are the Riemann, Weyl and Ricci tensors and the scalar curvature.
Term by term, we have:
\begin{eqnarray}
&C_{\mu \nu \lambda \sigma}C^{\mu \nu \lambda \sigma}
& = 48\left(\frac{1}{3}\frac{\partial M}{\partial R}\left(
\frac{\partial r}{\partial R}\right)^{-1}r^{-2}-Mr^{-3}\right)^2
\label{F11} \;,\\
&R_{\mu \nu}R^{\mu \nu}&= 64 \pi^2 \rho^2 = \left(2\frac{\partial M}{\partial R}\left(\frac{\partial r}{\partial R}\right)^{-1}r^{-2}\right)^2 \label{F12} \;, \\
&\CR& = 8\pi \rho = 2\frac{\partial M}{\partial R}\left(\frac{\partial r}{\partial R}\right)^{-1}r^{-2} \label{F13}\;.
\end{eqnarray}

For the separable solution \eqref{eq:34A-d2}, with $C_2 = 0$ corresponding to
the OS model (see, subsection \ref{FRW}), we get:
\begin{eqnarray}
   &C_{\mu \nu \lambda \sigma}C^{\mu \nu \lambda \sigma}& = 48\left(AM^{\frac{1}{2}}r^{-\frac{3}{2}}-Mr^{-3}\right)^2 = 0\label{OS1} \;,\\
  &R_{\mu \nu}R^{\mu \nu}&= 36A^2Mr^{-3} = 576\, \lambda ^4\, \epsilon^{-4} \label{kOS2} \;, \\
  &\CR& = 6AM^{\frac{1}{2}}r^{-\frac{3}{2}}= 24\, \lambda ^2\,\epsilon^{-2}  \label{kOS3} \;,\\
&K& = 108 A^2Mr^{-3}-96A M^{\frac{3}{2}}r^{-\frac{9}{2}}+48M^2r^{-6} = \frac{240}{81}\,\lambda ^4\, \epsilon^{-4}  \label{kOS4}\;,
  \end{eqnarray}
where $\epsilon := \lambda t + C_1$. The solution \eqref{eq:34A-d2} implies that  for $\epsilon \rightarrow 0$ we have
the gravitational singularity.

%%%%%%%%%%%%%%%%%%%%%%%%%%%%%%
\subsubsection{Apparent horizon}
%%%%%%%%%%%%%%%%%%%%%%$%%%%%%%

Very often, in the case of dynamical space-times it is not possible to establish
the event horizon (EH), since it requires a knowledge of the whole space-time,
including null and spatial infinities. One of the ways to replace the notion of
event horizon by something in principle determinable locally is to resort to the
outermost marginally outer trapped surface called the apparent horizon (AH).
Although its definition is dependent on the foliation of space-time, it is still
one of the most convenient ways to define the black hole, especially in the
numerical general relativity. In the spherically symmetric configuration, the AH
is a sphere residing on the hyper-surface.  Condition for this sphere to be an
outer trapped surface requires calculating the divergence of the outgoing (from
the surface) null vectors and demanding it to be zero. The radial null vector
associated with the LTB metric (for
$E=0$) in PG coordinates (see e.g. \cite{LLB} for derivation) reads:
\begin{equation}
\label{null-vector}
k^{\mu} = \left(1, 1-\sqrt{\frac{2M}{R}}, 0, 0\right) \;.
\end{equation}
The condition for the divergence of \eqref{null-vector} to be zero,
$k^{\mu}_{\;\; ;\mu} =
0$, on a given surface is equivalent to requiring that for this surface we have
\begin{equation}
\label{hor5}
r(t) = 2M(t,r(t)) \;.
\end{equation}
It is worth noting that for the surfaces located in the vacuum part of the
solution this condition is equivalent with the condition for the EH.

%%%%%%%%%%%%%%%%%%%%%%%%%%%%%%%%%%%%%%%%%%
\subsection{Oppenheimer-Snyder black hole}  \label{FRW}
%%%%%%%%%%%%%%%%%%%%%%%%%%%%%%%%%%%%%%%%%%

In the following we will focus on the interior region of the solution.

\noindent For $C_2 = 0$, we have from \eqref{eq:34A-d2} the formula
\begin{equation}\label{OS1}
  M(t,r) = \frac{2}{9}\,\frac{\lambda^2\,r^3}{(\lambda t + C_1)^2} \, ,
\end{equation}
which resolved with respect to $r$ gives
\begin{equation}\label{OS2}
  r = \Big(\frac{9}{2} \Big)^{1/3}M^{1/3} \Big(\frac{\lambda t +C_1}{\lambda} \Big)^{2/3} \, .
\end{equation}

Eq. \!\eqref{OS2} shows  that specifying $M$ and $t$ leads to corresponding $r$. Thus, inserting into \eqref{OS2}
$M = M(0, R) = M_0$ leads to
\begin{equation}\label{OS3}
  r = r_0 (t,R) =: f(R) g(t) \, .
\end{equation}

The covariant condition for spherically symmetric dust solution of Einstein
equations to be equivalent to the Friedmann–Lema\^{i}tre–Robertson–Walker (FLRW) model is vanishing of the shear, acceleration and rotation (see,
e.g., \cite{PlebanskiKrasinski}).  The only non-trivial quantity in our case  is the shear $\sigma$, which reads
\begin{equation}
\label{eq:ShearSigma}
\sigma = \frac{1}{3}\left(\frac{1}{r}\frac{\partial r}{\partial t}
-\left(\frac{\partial r}{\partial R}\right)^{-1}
\frac{\partial^2 r}{\partial t \partial R}\right)\;.
\end{equation}

Taking $r = f(R) g(t)$ gives $\sigma = 0$ and hence $C_2=0$ leads to the FLRW model and after imposing
the Schwarzschild solution in the outer region, to the  OS
black hole.  This can be achieved by prescribing the initial density profile, e.g., with the use of
the Heaviside step function. Such setup guaranties the Schwarzschild solution in the exterior region
throughout the evolution.

The equation defining the horizon can be written, due to \eqref{hor5}, in the form
\begin{equation}\label{OS3}
  r_0 (t_h,R) = 2 M_0  \, ,
\end{equation}
where $t_h$ denotes the time at which the most outer shell reaches the horizon.
Resolving \eqref{OS3} with respect to $t_h$ leads to the formula $ t_h=t_c \mp \frac{4}{3} M_0 $.

Since we consider the collapsing star, the comoving observer passes the horizon before approaching the gravitational
singularity. At the singularity the classical dynamics breaks down. Thus, there is no classical evolution
for $t \geq t_c$ so that we choose
\begin{equation}\label{OS4}
 t_h=t_c - \frac{4}{3} M_0 \, .
\end{equation}
%

%%%%%%%%%%%%%%%%%%%%%%%%%%%%%%%%%%%%%%%%%%%%%%%
\section{Quantum Oppenheimer-Snyder black hole} \label{QOS}
%%%%%%%%%%%%%%%%%%%%%%%%%%%%%%%%%%%%%%%%%%%%%%%

In the standard approach to quantization of classical dynamics defined by the
Hamiltonian $H$, that is a generator of the dynamics, one maps $H$ onto an
operator $\hat{H}$ defined in some Hilbert space $\mathcal{H}$. If $\hat{H}$ is
self-adjoint in $\mathcal{H}$, it can be used to define quantum dynamics in the
form of the Schr\"{o}dinger equation defined in $\mathcal{H}$.

In the case considered in this paper, we do not quantize Hamilton's
dynamics, but we use the solution to the classical dynamics to construct three
quantum observables generated by their classical counterparts: the time operator
$\hat{t}$, the radius operator $\hat{r}$ and the total mass operator
$\hat{M}$.  To this purpose we use the classical dynamics similarly as
in the cosmological case considered in \cite{bkl}. The classical dynamics is
defined by the equations of motion obtained by inserting the metric \eqref{d2}
into the Einstein equations. In what follows we consider the marginally bound
model so that the dynamics is defined by Eq. \!\eqref{dd5}.  The equations of
motion, with suitable set of boundary condition, carry the same information as
the general solution to these equations. In our case as
Eq. \!\eqref{eq:34A-d2} presents the solution to Eq. \!\eqref{dd5} so that the
solution $M(t,r)$ includes the dynamics of this gravitational system.

%%%%%%%%%%%%%%%%%%%%%%%%%%%%%%%%%%%%%%%%%
\subsection{Integral quantization method}
%%%%%%%%%%%%%%%%%%%%%%%%%%%%%%%%%%%%%%%%%

In what follows we apply the affine coherent states (ACS) quantization (see,
\cite{Ola,AWG,AWT,AW,bkl} and references therein) to the quantization of the classical
dynamics \eqref{dd5} corresponding to the marginally bound case. The extension
to the general dust case \!\eqref{dd3}--\eqref{dd4} and perfect fluid case can
be done by analogy.

We begin with introducing the extended configuration space  $T$ for our system. It is
defined as follows \cite{Ola}
\begin{equation}\label{t1}
  T = \{(t,r) \in \dR \times \dR_+\},~~~~\dR_+ = (0, +\infty) \, ,
\end{equation}
where $t$ and $r$ are the time and radial coordinates, respectively, which occur
in the line element \eqref{d2}.

Since the configuration space is a half plane, it can be identified with the
affine group $\Aff =:G$ for which the multiplication rule is given by

\begin{equation}\label{c1b}
(t^\prime, r^\prime)\cdot (t, r) = (r^\prime t + t^\prime, r^\prime r ),
\end{equation}
with the unity $(0,1)$ and the inverse
\begin{equation}\label{c2b}
(t, r)^{-1} = \left(-\frac{t}{r}, \frac{1}{r}\right).
\end{equation}

The affine group has two, nontrivial, inequivalent  irreducible unitary
representations.  Both are realized in the Hilbert space
$\mathcal{H}=L^2(\dR_+, d\nu(x))$, where $d\nu(x)=dx/x$ is the invariant
measure on the multiplicative group $(\dR_+,\cdot)$.
In what follows we choose the one defined by
\begin{equation}\label{im1b}
U(t,r)\psi(x)= e^{i t x} \psi(rx)\, .
\end{equation}
The integration over the affine group reads
\begin{equation}\label{m5}
\int_G d\mu (t,r)
:= \frac{1}{2\pi}\int_{-\infty}^\infty dt \int_0^\infty dr /r^2\, ,
\end{equation}
where the measure $d\mu (t,r)$ is left invariant.

Fixing the normalized vector $\Ket{\Phi_0} \in L^2(\dR_+, d\nu(x))$, called the
fiducial vector, one can define a continuous family of affine coherent states
$\Ket{t,r} \in L^2(\dR_+, d\nu(x))$ as follows
\begin{equation}\label{im2}
\Ket{t,r} = U(t,r) \Ket{\Phi_0}.
\end{equation}
The fiducial vector can be
taken to be any vector of $\mathcal{H}$ that satisfies certain conditions to
be specified  during the quantization process.  It is a sort of free
``parameter'' of  ACS quantization. First of all, it should be normalized so
that we should have
\begin{equation}\label{fid}
  \BraKet{\Phi_0}{\Phi_0} := \int_0^{\infty} d\nu (x)
  \BraKet{\Phi_0}{x} \BraKet{x}{\Phi_0}
  =  \int_0^{\infty} d\nu (x) |\Phi_0(x)|^2 = 1 \, ,
\end{equation}
where we have used the formula  \cite{AWT}
\begin{equation}\label{xxx}
\int_0^{\infty} d\nu (x) \Ket{x}\Bra{x}  =  \UnitOp \, ,
\end{equation}
which applies to $\mathcal{H}$.

The space of coherent states is highly entangled in the sense that  we have
\begin{equation}\label{ent1}
  \langle t,r |t^\prime,r^\prime \rangle \neq 0~~~ \mbox{if}~~~t \neq t^\prime~~\mbox{or}~~ r \neq r^\prime \, ,
\end{equation}
\begin{equation}\label{ent2}
\langle t,r | t,r \rangle = 1~~~ \mbox{if}~~~\langle \Phi_0|\Phi_0 \rangle = 1 \, .
\end{equation}

The  irreducibility of the representation, used to define the coherent
states \eqref{im2}, enables making use of Schur's lemma, which leads
to the  resolution of the unity $\UnitOp$  in $L^2(\dR_+, d\nu(x))$:
\begin{equation}\label{im4}
\int_{G}  d\mu(t,r) \Ket{t,r}\Bra{t,r} = A_{\Phi_0}\; \UnitOp\; ,
\end{equation}
where
\begin{equation}\label{im3b}
A_{\Phi_0} := \int_0^\infty \frac{dp}{p^2} |\Phi_0(p)|^2 < \infty\, .
\end{equation}

Making use of the resolution of the unity \eqref{im4}, we define the
quantization of a  classical observable $f: T \rightarrow \dR$ as follows
\begin{equation}\label{im8}
 f \longrightarrow  \hat{f} :=
\frac{1}{A_{\Phi_0}}\int_{\Group{G}} d\mu(t,r)
\Ket{t,r} f(t,r) \Bra{t,r}  \, ,
\end{equation}
where $\hat{f}: \mathcal{H} \rightarrow \mathcal{H}$ is the corresponding
quantum observable.

The mapping \eqref{im8} is  covariant in the sense that one has
\begin{eqnarray}\label{cov}
&& U(\xi_0) \hat{f} U^\dag (\xi_0)
= \frac{1}{A_{\Phi_0}}\int_{\Group{G}} d\mu(\xi)
U(\xi_0)\Ket{\xi} f(\xi) \Bra{\xi}U^\dag (\xi_0) \nonumber \\
&& =\frac{1}{A_{\Phi_0}}\int_{\Group{G}} d\mu(\xi)
\Ket{\xi} f(\xi_0^{-1}\cdot \xi) \Bra{\xi} \, ,
\end{eqnarray}
where
$\xi_0^{-1}\cdot \xi= (t_0,r_0)^{-1}\cdot (t,r) =
(\frac{t-t_0}{r_0},\frac{r}{r_0})$, with $\xi := (t,r)$.  It means, no point in
the configuration space $T$ is privileged.

Eq. \!\eqref{im8} defines a linear mapping and the observable $\hat{f}$ is a
symmetric operator by the construction. It results from the Schwartz inequality that
$|\BraKet{t,r}{\Psi}| \leq \Vert \Psi \Vert$ for every vector
$\Psi \in \mathcal{H}$ and every $(t,r)$. This implies that the sesquilinear
form corresponding to the operator $\hat{f}$ fulfill the following inequalities
\begin{eqnarray}
\label{OperNormQuantOper}
&&|\Bra{\Psi_2} \hat{f}\Ket{\Psi_1}|
\leq \frac{1}{A_\phi} \int_{\Group{G}} d\mu(t,r) |f(t,r)|
|\BraKet{t,r}{\Psi_2}^\star||\BraKet{t,r}{\Psi_1}| \nonumber \\
&&\leq \left[ \frac{1}{A_\phi} \int_{\Group{G}} d\mu(t,r) |f(t,r)| \right]
\Vert \Psi_2 \Vert  \Vert \Psi_1 \Vert \, .
\end{eqnarray}
The second inequality proves that for every finite value of the integral in the
square bracket, i.e., $\int_{\Group{G}} d\mu(t,r) |f(t,r)| < \infty$, the
operator $\hat{f}$ is bounded so that it is a self-adjoint operator.
% W.~Mlak, Wstęp do teorii przestrzeni Hilberta, Biblioteka Matematyczna
% vol. 35, PWN, Warszawa 1972, p.143.

%%%%%%%%%%%%%%%%%%%%%%%%%%%%%
\subsection{Quantum dynamics}
\label{sec:QuantumDynamics}
%%%%%%%%%%%%%%%%%%%%%%%%%%%%%

We propose to apply the procedure of mapping a classical dynamics onto a quantum
dynamics considered in \cite{bkl}.  In that method, the time variable of the
classical level is mapped onto an operator acting at the quantum level so that
it is no longer an evolution parameter of the quantum level, but a quantum
observable like other quantum observables.  In this way, the time is considered
on the same footing as the spatial coordinates.

In our classical model we have three observables: the time-like $t$, the
space-like $r$, and the mass $M(t,r)$. The maximum $M(t,r)$ equal to $M_0$,
represents the total energy of this system which has to be a conserved quantity.
The imposition of the energy conservation constraint onto the mass operator
$\hat{M}$, calculated in the appropriate Hilbert space, leads to a quantum
dynamics.

Classical observables should be related to the corresponding quantum observables
by their expectation values.  This is the leading idea we use to find quantum
states of our system in a Hilbert space. In our case, these three classical
basic observables are mapped onto the corresponding operators $\hat{t}$,
$\hat{r}$, and $\hat{M}$ by using Eq. \!\eqref{im8}.
%The cesponding operators $\hat{r}$, $\hat{t}$ and $\hat{M}$ by using
%Eq. \!\eqref{im8}.
Since $M$ is of a fundamental importance, as it represents the global energy of
considered gravitational system, we treat the corresponding operator $\hat{M}$
as the main quantum observable. We use the expectation value of $\hat{M}$ to
determine the family of vector states $\psi_\eta (t,r)$ in the Hilbert space
$\StateSpace{K}= L^2(G,d\mu(t,r))$, where $\eta = (\eta_1,\eta_2)$ are the
parameters specifying the quantum evolution.

Consistently, we require the states $\Ket{\psi_\eta}$ to satisfy the constraint
\begin{equation}\label{cons1}
\Bra{\psi_\eta} \hat{M} \Ket{\psi_\eta} = M_0 \, ,
\end{equation}
where $M_0$ is the mass of considered dust star defined by the density $\rho$
of the  star with the radius $r_0$. For the OS star with the density
\begin{equation}\label{den1}
\rho (t,r) =\left\{
\begin{array}{lll}
\frac{\lambda^2}{6\pi}\frac{1}{(\lambda t+ C_1)^2}&\mbox{for}& r \leq r_0 (t)\\
0&\mbox{for}&r > r_0 (t)
\end{array}
\right.\; ,
\end{equation}
the corresponding mass defined by Eq. \!\eqref{eq6} reads
\begin{equation}\label{mass1}
M (t,r) =\left\{
\begin{array}{lll}
\frac{2\lambda^2}{9}\frac{r^3}{(\lambda t+ C_1)^2}&\mbox{for}& r \leq r_0 (t)\\
M_0&\mbox{for}&r > r_0 (t)
\end{array}
\right.\; ,
\end{equation}
so that the relationship between $M_0$ and $r_0$ is
\begin{equation}\label{bet}
r_0 (t)
= \Big(\frac{9 M_0}{2 \lambda^2} \Big)^{1/3}(\lambda t + C_1)^{2/3}
\, .
\end{equation}
Due to \eqref{mass1}, the expression for the mass operator can be rewritten as
\begin{equation}
\label{eq:MassOpOS}
\hat{M} = \frac{1}{2\pi A_\phi} \int_{-\infty}^{\infty} dt \, \left\{
\int_{0}^{r_0(t)} \frac{dr}{r^2}
\Ket{t,r}
\left[\frac{2\lambda^2}{9}\frac{r^3}{(\lambda t+ C_1)^2} \right]
\Bra{t,r}
+ M_0 \int_{r_0(t)}^{\infty} \frac{dr}{r^2} \Ket{t,r} \Bra{t,r} \right\} \, .
\end{equation}
This operator fulfill the condition \eqref{OperNormQuantOper}
\begin{equation}
\label{eq:MassOpBound}
|\Bra{\Psi_2} \hat{M} \Ket{\Psi_1}| \leq
M_0 \Vert \Psi_2 \Vert  \Vert \Psi_1 \Vert \, ,
\end{equation}
which implies that $\hat{M}$ is a bounded operator.

Using the resolution of the unity \eqref{im4}, the mass operator can be
rewritten as
\begin{equation}
\label{eq:2MassOpOS}
\hat{M} = M_0 \UnitOp + \frac{1}{2\pi A_\phi} \int_{-\infty}^{\infty} dt \,
\int_{0}^{r_0(t)} \frac{dr}{r^2}
\Ket{t,r}
\left[\frac{2\lambda^2}{9}\frac{r^3}{(\lambda t+ C_1)^2} -M_0 \right]
\Bra{t,r}  \, .
\end{equation}
This form of the mass operator shows that the calculated mass is independent of
the shape of the wave function $\Psi(t,r):=\BraKet{t,r}{\Psi}$ outside of the
star, i.e., for $r \ge r_0(t)$. This also implies that all states
$\Ket{\Psi_{M_0}} \in \StateSpace{H}$ satisfying the condition
$\BraKet{t,r}{\Psi_{M_0}}=0$ for all $r \leq r_0(t)$ are eigenstates of the mass
operator which belong to the eigenvalue $M_0$
\begin{equation}
\label{eq:EigenVecM0}
\hat{M} \Ket{\Psi_{M_0}}= M_0 \Ket{\Psi_{M_0}} \, ,
\end{equation}
i.e., $M_0$ is the upper bound of the mass spectrum. On the other hand, the mass
function is non-negative. The condition $M(t,r) \ge 0$ determines the lower
bound of the mass operator spectrum to be equal to zero.

According to the interpretation of the state spaces of our quantum system
described in App.~\eqref{app:StateSpaces}, the corresponding scalar product
$\BraKet{t,r}{\Psi_{M_0}}$ should be interpreted as the probability amplitude of
finding our OS star at the time $t$ in a form of a spherical three dimensional
object with the radius $r$. The constructed mass operator $\hat{M}$ (see
\eqref{eq:MassOpOS}), measures the full mass of the OS star bounded by the
classical radius $r_0(t)$. This feature and the fact that $M_0$ is the maximum
of the mass operator, imply that all the amplitudes $\BraKet{t,r}{\Psi}$ have to
be equal to zero inside the star, for every acceptable physical state
$\Ket{\Psi}=\Ket{\Psi_{M_0}} \in \StateSpace{H}$.

The above suggests that the mass operator \eqref{eq:MassOpOS} has also
eigenstates corresponding to masses smaller than $M_0$. It is clear that all the
states $\Ket{\Psi} \in \StateSpace{H}$ for which the amplitudes
$\BraKet{t,r}{\Psi}$ are not identically equal to zero inside of the star
correspond to smaller masses than $M_0$.  In the case of an isolated OS star,
such states do not occur because of the conservation of the total
energy. However, some perturbations of the eigenstates \eqref{eq:EigenVecM0} can
lead to the mass smaller than $M_0$ and possibly a new dynamics of the OS
star.
%%%%%%%%%%%%%%%

Such mass variation may come, for example,  from the physical process of the
Hawking radiation. However, some analyses suggest (see, e.g., \cite{Hal}) that this
dissipative process has minor effect on deeply quantum regime inside black hole
(but  should be implemented into analysis for completeness).
Our formalism gives some possibility of a description
of such a process. However, mathematical description requires the preparation
of a kind of transition operator which transforms a state of given
mass to a state of lower mass.
This interesting feature requires further analysis and will be consider
elsewhere.

In what follows, we consider the quantum states corresponding to the total
mass $M_0$, defined to be wave packets localized within a given time and space
intervals, and having additionally a ``hole'' including the singularity at
$t=t_c$.  Qualitatively, each packet represents a wave function with compact
support.  The wave function without a hole can be obtained in the limit
$\epsilon \to 0^+$.  This type of wave packet can be written as
\begin{equation}
\label{soph5}
\psi_{\eta_1,\eta_2}(t,r)
=\frac{1}{\sqrt{N}}\,r\,\chi_{\left(r(t,\eta_1),r(t,\eta_1+\delta_1)
  \right)}(r)\,
\chi_{\left(\eta_2-\frac{\delta_2}{2},\eta_2+\frac{\delta_2}{2}\right)}(t)\,
\left(1-\chi_{\left(t_c-\frac{\epsilon}{2},t_c
 +\frac{\epsilon}{2}\right)}(t)\right) \, ,
\end{equation}
where $\delta_1,\delta_2>0$, $\eta_1\in \langle 0,\infty)$,
$\eta_2 \in (-\infty,\infty)$. In addition we assume the condition
$0 < \epsilon < \delta_2$, which defines the wave functions with the width larger
than the hole around the singularity. It means, we consider a subset of wave
functions which totally cover the hole. They are equal to zero within the hole
and are different from zero outside the hole. In this way they connect both
regions separated by the singularity.

In what follows we use the following auxiliary functions:
The characteristic function $\chi_{(a,b)}(x)$ of the set $(a,b)$ defined as
\begin{equation}
\chi_{(a,b)}(x)=\left\{
\begin{array}{lll}
1&\mbox{for}&x\in(a,b)\\
0&\mbox{for}&x\not\in (a,b)
\end{array}
\right.\; ,
\end{equation}
and the function $r(x,y)$ defined to be
\begin{equation}
r(x,y)=\sqrt[3]{\frac{9}{2}}y^{\frac{1}{3}}(x-t_c)^{\frac{2}{3}}\; ,
\qquad t_c=-\frac{C_1}{\lambda} \, ,
\end{equation}
which  represents the radius of the shell with the mass $y$ at the time $x$.

Next, we define two new functions in terms of integrals
\begin{equation}
\label{eq:GcalFun}
\mathcal{G}_\alpha(a,b)
=\int_a^b dx\, x^{\alpha-1}
=\frac{1}{\alpha}\left(b^\alpha-a^\alpha\right)\, ,\quad\alpha\neq 0 \, ,
\end{equation}
and
\begin{eqnarray}
\label{eq:FcalFun}
&&\mathcal{F}_\alpha(\eta_2)=
\int_{\eta_2-\frac{\delta_2}{2}}^{\eta_2+\frac{\delta_2}{2}}dx\,
(x-t_c)^{\alpha-1}
\left(
1-\chi_{\left(t_c-\frac{\epsilon}{2},\;t_c+\frac{\epsilon}{2}\right)}(x)
\right)
\\
&&=
\left\{
\begin{array}{lll}
\mathcal{G}_\alpha\left(\eta_2^-,\eta_2^+\right)
&\mbox{for}& \eta_2 \leq t_c-\frac{\epsilon+\delta_2}{2}
\\
\mathcal{G}_\alpha\left(\eta_2^-,-\frac{\epsilon}{2}\right)
&\mbox{for}& t_c-\frac{\epsilon+\delta_2}{2} < \eta_2
\leq t_c-\frac{\delta_2-\epsilon}{2}
\\
\mathcal{G}_\alpha\left(\eta_2^-,-\frac{\epsilon}{2}\right)+
\mathcal{G}_\alpha\left(\frac{\epsilon}{2},\eta_2^+\right)
&\mbox{for}& t_c-\frac{\delta_2-\epsilon}{2} < \eta_2
\leq t_c+\frac{\delta_2-\epsilon}{2}
\\
\mathcal{G}_\alpha\left(\frac{\epsilon}{2},\eta_2^+\right)
&\mbox{for}& t_c+\frac{\delta_2-\epsilon}{2}< \eta_2
\leq t_c+\frac{\delta_2+\epsilon}{2}
\\
\mathcal{G}_\alpha\left(\eta_2^-,\eta_2^+\right)
&\mbox{for}& \eta_2 > t_c+\frac{\delta_2+\epsilon}{2}
\end{array}
\right.
\end{eqnarray}
where
\begin{equation}
\eta_2^- :=\eta_2-\frac{\delta_2}{2}-t_c,
\qquad
\eta_2^+ :=\eta_2+\frac{\delta_2}{2}-t_c~\, .\\
\end{equation}

It is worth to notice that we have
\begin{equation}
\label{eq:FLim}
\mathcal{F}_{\alpha}(\eta_2)\xrightarrow[]{\epsilon\rightarrow 0}\left\{
\begin{array}{lll}
\mathcal{G}_\alpha(\eta_2^-,\eta_2^+)&\mbox{for}&\alpha>0\\
\infty&\mbox{for}& \alpha< 0 \, .
\end{array}
\right.
\end{equation}

The normalization constant of the wave packets \eqref{soph5}  reads
\begin{equation}
N=\frac{1}{2\pi A_\phi}\frac{1}{3}\,\sqrt[3]{\frac{9}{2}}\;
\mathcal{G}_{\frac{1}{3}}(\eta_1,\eta_1+\delta_1)\;
\mathcal{F}_{\frac{5}{3}}(\eta_2) \, .
\end{equation}
The expectation value of the operator $\hat{t}$ is the following
\begin{eqnarray}
&& t_Q(\eta_1,\eta_2):=
\Bra{\psi_{\eta_1,\eta_2}}
\hat{t}
\Ket{\psi_{\eta_1,\eta_2}} \nonumber\\
&& =\frac{1}{2\pi A_\phi}\frac{1}{N}
\int_{\eta_2-\frac{\delta_2}{2}}^{\eta_2+\frac{\delta_2}{2}}dt\, t
\left(1-\chi_{
  \left(t_c-\frac{\epsilon}{2},t_c+\frac{\epsilon}{2}\right)}(t)
\right)
\int_{r(t,\eta_1)}^{r(t,\eta_1+\delta_1)}dr \nonumber\\
&& =\frac{\mathcal{F}_{\frac{8}{3}}(\eta_2)}{\mathcal{F}_{\frac{5}{3}}(\eta_2)}
+t_c \label{czas} \,.
\end{eqnarray}
The expectation value of the operator $\hat{r}$ is found to be
\begin{eqnarray}
&& r_Q(\eta_1,\eta_2):=
\Bra{\psi_{\eta_,\eta_2}}
\hat{r}
\Ket{\psi_{\eta_1,\eta_2}} \nonumber \\
&& =\frac{1}{2\pi A_\phi}\frac{1}{N}
\int_{\eta_2-\frac{\delta_2}{2}}^{\eta_2+\frac{\delta_2}{2}}dt
\left(1-\chi_{\left(t_c-\frac{\epsilon}{2},t_c+\frac{\epsilon}{2}\right)}(t)
\right)\,
\int_{r(t,\eta_1)}^{r(t,\eta_1+\delta_1)}dr\, r \nonumber\\
&& =\sqrt[3]{\frac{9}{2}}\;
\frac{\mathcal{G}_{\frac{2}{3}}(\eta_1,\eta_1+\delta_1)}{
\mathcal{G}_{\frac{1}{3}}(\eta_1,\eta_1+\delta_1)}\;
\frac{\mathcal{F}_{\frac{7}{3}}(\eta_2)}{\mathcal{F}_{\frac{5}{3}}(\eta_2)}
\label{pro}\,.
\end{eqnarray}
The expectation value of the mass operator $\hat{M}$ reads
\begin{equation}
\label{eq:AverMOp}
\Bra{\psi_{\eta_1,\eta_2}} \hat{M} \Ket{\psi_{\eta_1,\eta_2}} = M_0 \, ,
\end{equation}
where $\eta_1 \geq M_0$.

For very small values of $\delta_2$ and $\epsilon$ the functions
$t_Q(\eta_1,\eta_2)$ and $r_Q(\eta_1,\eta_2)$ can be approximated as follows
\begin{eqnarray}
&& \label{smallt} t_Q(\eta_1,\eta_2) \approx \eta_2 \, ,\\
&& \label{smallr} _Q(\eta_1,\eta_2) \approx
\sqrt[3]{\frac{9}{2}} \eta_1^{\frac{1}{3}} (\eta_2-t_c)^{\frac{2}{3}} \, .
\end{eqnarray}
Comparing these formulae with their classical counterparts, we get
$\eta_1 = M_0$,  and we see that the parameter $\eta_2$ corresponds to the  classical time
(but only to some extent).

\begin{figure}[]
\includegraphics[height=5cm]{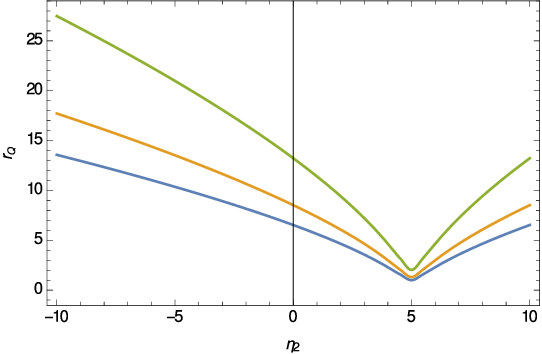}
\caption{The $r_{Q}(\eta_1,\eta_2)$ function (see Eq. \!\eqref{pro}) for $\eta_2 \in [-10,10]$,
with $\eta_1=2=M_0$ (blue curve), $\eta_1=5$ (orange curve), $\eta_1=20$ (green curve),
 where
$\delta_1=\delta_2=1$, $\epsilon=10^{-5}$, $t_c=5$.}
\label{Plot:rQ}
\end{figure}

With fixed $\eta_1$, the quantum radius of the OS star $r_Q(\eta_1,\eta_2)$
is found to achieve its minimum at $\eta_2=t_c$ and reads
\begin{equation}
\label{eq:MinrQ}
r_{Q}(\eta_1,t_c)=
\sqrt[3]{\frac{9}{2}}\;
\frac{ \mathcal{G}_{\frac{2}{3}}(\eta_1,\eta_1+\delta_1)}{
\mathcal{G}_{\frac{1}{3}}(\eta_1,\eta_1+\delta_1)}\;
\frac{5}{7} \left(\frac{1}{2}\right)^{\frac{2}{3}}
\frac{
\delta_2^{\frac{7}{3}}-\epsilon^{\frac{7}{3}} }{
\delta_2^{\frac{5}{3}}-\epsilon^{\frac{5}{3}}
} \, .
\end{equation}
The radius $r_{Q}(\eta_1,t_c)$ represents the bouncing radius of the OS
star.
The corresponding quantum time $t_{Q}(\eta_1,t_c)=t_c$,
as $\mathcal{F}_{\frac{8}{3}}(t_c)=0$ in \eqref{czas}, so that it is equal to the
time at which the classical singularity is achieved.   After the bounce the quantum star expands (see Figure \ref{Plot:rQ}).
This phenomenon does not disappear even for the states without the whole, i.e., for $\epsilon=0$.

Another interesting aspect of the evolution of the OS star is its horizon, which
the collapsing star reaches, corresponding to the classical time $t_h$ given by
the formula \eqref{hor5}.  To estimate the parameters $\eta_2$ , while
$\eta_1 = M_0$, let us impose the condition
\begin{equation}
\label{eq:QuanTHorizon}
t_Q(M_0,\eta_2)= t_h = t_c - \frac{4}{3} M_0.
\end{equation}
For small $\delta_1,\delta_2$ and $\epsilon$, the approximate solution of
Eq.~\eqref{eq:QuanTHorizon} is found to be $\eta_2 = t_c - \frac{4}{3} M_0$. In
this approximation the horizon radius $r_Q(M_0,\eta_{2}) \approx 2M_0$, i.e., it
reproduces the classical condition for the horizon.

The vector state in the form of the wave packet \eqref{soph5} is well defined
for any $t$, including the neighborhood of the singularity at $t_c$. Thus,
the quantum dynamics in terms of expectation values \eqref{czas} and \eqref{pro}
is well defined before and after the singularity. See, for instance, the case
of a very small $\delta_2$ and $\epsilon$ presented by \eqref{smallt} and \eqref{smallr}.
In particular, the equation
\begin{equation}\label{two-hor}
 \Bra{\psi_{\eta_1,\eta_2}} \hat{r}
\Ket{\psi_{\eta_1,\eta_2}}  = 2\; \Bra{\psi_{\eta_1,\eta_2}} \hat{M} \Ket{\psi_{\eta_1,\eta_2}} \, ,
\end{equation}
for $\eta_1 = M_0$, is formally well defined. This equation corresponds to the classical equation \eqref{hor5}
that determines the horizon \eqref{OS4}, but is defined at the quantum level. Due to \eqref{pro} and \eqref{eq:AverMOp},
Eq.~\eqref{two-hor} can be rewritten as
\begin{equation}\label{2-hor}
 \sqrt[3]{\frac{9}{2}}\;
\frac{\mathcal{G}_{\frac{2}{3}}(M_0,M_0+\delta_1)}{
\mathcal{G}_{\frac{1}{3}}(M_0,M_0+\delta_1)}\;
\frac{\mathcal{F}_{\frac{7}{3}}(\eta_2)}{\mathcal{F}_{\frac{5}{3}}(\eta_2)}
 = 2 M_0 \, .
\end{equation}

Let us solve numerically Eq.~\eqref{2-hor}. We take $t_c = 1$ and $M_0=\eta_1=1$; the
state parameters $\delta_1=0.01$, $\delta_2=0.2$, and $\epsilon=0.1$. For such values of the state
parameters Eq.~\eqref{two-hor} has two solutions for $\eta_2$, which we denote as
$\eta_{2h} = 1 \mp 1.32877 \approx t_c \mp \frac{4}{3}M_0$.
The second solution, $\eta_{2h} = 1 + 1.32877$,  that follows the bounce does not have as
clear physical meaning as the one preceding the bounce. It will be further discussed in Sec. \ref{Con}.

These numerical results are not changed qualitatively for different choice of parameters
and also for the states without a hole, i.e., for $\epsilon=0$.

%%%%%%%%%%%%%%%%%%%%%%%%%%%%%%%%%%%%%%%%%%%%%%%%%%%%%%%
\subsection{Fate of classical singularities} \label{QOS}
%%%%%%%%%%%%%%%%%%%%%%%%%%%%%%%%%%%%%%%%%%%%%%%%%%%%%%%

In what follows we address the issue of the classical singularities of the
Oppenheimer-Snyder black hole at the quantum level.

Let us introduce the following function
\begin{equation}
\mathcal{A}(\alpha,\beta,\gamma)=\frac{1}{(t-t_c)^\alpha}M(t,r)^\beta r^{-\gamma} \,.
\end{equation}
Our curvature invariants \eqref{kOS2}-\eqref{kOS4} can be expressed in terms of
$\mathcal{A}$ in the following way:
\begin{eqnarray}
&&\label{inv-1} R_{\mu\nu}R^{\mu\nu}=8\,\mathcal{A}(2,1,3) \, ,\\
&&\label{inv-2} \mathcal{R}=(-2\sqrt{2})\,\mathcal{A}\left(1,\frac{1}{2},\frac{3}{2}\right) \, ,\\
&&\label{inv-3} K=48\,\mathcal{A}(2,1,3)+32\sqrt{2}\,\mathcal{A}\left(2,\frac{3}{2},\frac{9}{2}\right)+48\,\mathcal{A}(0,2,6) \, .
\end{eqnarray}

The expectation value of the operator $\hat{\mathcal{A}}(\alpha,\beta,\gamma)$
is found to be:
\begin{eqnarray}
\label{eq:ExpAOp}
&&\Bra{\psi_{\eta_1,\eta_2}}
\hat{\mathcal{A}}(\alpha,\beta,\gamma)
\Ket{\psi_{\eta_1,\eta_2}} \nonumber\\
&&
=\frac{1}{2\pi A_\psi}\, \frac{1}{N}
\int_{\eta_2-\frac{\delta_2}{2}}^{\eta_2+\frac{\delta_2}{2}}dt
\left(1-\chi_{\left(t_c-\frac{\epsilon}{2},t_c
  +\frac{\epsilon}{2}\right)}(t)\right)\,
\int_{r(t,\eta_1)}^{r(t,\eta_1+\delta_1)}dr \frac{M(t,r)^\beta}{(t-t_c)^\alpha
r^\gamma}   \nonumber \\
&&
=\left(\frac{2}{9}\right)^{\frac{\gamma}{3}}
\frac{\mathcal{F}_{\frac{5}{3}-\frac{2}{3}\gamma-\alpha}(\eta_2)}{
 \mathcal{F}_{\frac{5}{3}}(\eta_2)}\cdot  \\
&&
\cdot  \nonumber
\left\{
\begin{array}{lll}
\frac{\mathcal{G}_{\beta-\frac{\gamma}{3}+\frac{1}{3}}(\eta_1,\eta_1+\delta_1)}{
  \mathcal{G}_{\frac{1}{3}}(\eta_1,\eta_1+\delta_1)}&\mbox{for}&
\eta_1<M_0-\delta_1  \\
\frac{
\mathcal{G}_{\beta-\frac{\gamma}{3}+\frac{1}{3}}(\eta_1,M_0)+
M_0^\beta\mathcal{G}_{\frac{1}{3}-\frac{\gamma}{3}}(M_0,\eta_1+\delta_1)
}{\mathcal{G}_{\frac{1}{3}}(\eta_1,\eta_1+\delta_1)}&
\mbox{for} & M_0-\delta_1 \leq \eta_1 < M_0 \, . \nonumber \\
M_0^\beta\frac{\mathcal{G}_{\frac{1}{3}
-\frac{\gamma}{3}}(\eta_1,\eta_1+\delta_1)}{
\mathcal{G}_{\frac{1}{3}}(\eta_1,\eta_1+\delta_1)}&
\mbox{for} & \eta_1 \geq M_0
\end{array}
\right.
\end{eqnarray}

Is it possible for the expectation value of the operator $\hat{\mathcal{A}}$ to be infinite?
Let us examine this issue:

\noindent Since the numerators in Eq. \! \eqref{eq:ExpAOp} are defined in terms of the
functions $\mathcal{G}_\alpha(a,b)$, the infinity can be achieved if and only if
one has division by zero. In the denominators we get two kinds of functions
which potentially can be equal to zero. The first one is
$\mathcal{G}_{\frac{1}{3}}(a,b)=0$ for $a=b$. This condition cannot be satisfied
for $\delta_1>0$.  The second one is the function
$\mathcal{F}_{\frac{5}{3}}(\eta_2)$ which is a combination of the functions
$\mathcal{G}$.  Making use of the definition \eqref{eq:FcalFun} and the fact
that $\mathcal{G}_{\frac{5}{3}}(a,b)=0$ is true only for $a=b$, the only case of
division by zero is when $\eta_2^- \leq-\frac{\epsilon}{2}$ and
$\frac{\epsilon}{2} \leq \eta_2^+$. However, in the case $\delta_2>\epsilon$ the
two boundary conditions $\eta_2^- =-\frac{\epsilon}{2}$ and
$\frac{\epsilon}{2} =\eta_2^+$ can not be fulfilled simultaneously. Therefore,
the function $\mathcal{F}_{\frac{5}{3}}(\eta_2)$ is not equal to zero for any
$\eta_2$.
We conclude that the expectation value of the operator $\hat{\mathcal{A}}$ for the
states $\psi_{\eta_1\eta_2}$ is not singular. Therefore, the expectation values
of the operators corresponding to the invariants \eqref{inv-1}-\eqref{inv-3} nowhere diverge.

On the other hand, due to the equation \eqref{eq:FLim}, it is easily seen that we have
\begin{equation}\label{diver}
\Bra{\psi_{\eta_1,\eta_2}}\hat{\mathcal{A}}\Ket{\psi_{\eta_1,\eta_2}}\xrightarrow[]
{\epsilon\rightarrow 0}\infty \, ,
\end{equation}
so that the expectation values of the operators corresponding to \eqref{inv-1}--\eqref{inv-3} diverge.

It is instructive to examine the variance of this operator in the case $\epsilon\rightarrow 0^+$.
The variance of the symmetric operator $\hat{\mathcal{A}}$ in the quantum state $|\psi\rangle \in \mathcal{K}$
with compact support, for $\epsilon > 0$, is the following  \cite{Ola,Rob}
\begin{equation}\label{varA}
\Var{\hat{\mathcal{A}},\psi} := \langle \psi|\big(\hat{\mathcal{A}}-
\langle \hat{\mathcal{A}}\rangle_\psi\UnitOp  \big)^2 \psi\rangle =
\langle \big(\hat{\mathcal{A}}- \langle\hat{\mathcal{A}}\rangle_\psi\UnitOp \big)\psi |\big(\hat{\mathcal{A}}-
\langle\hat{\mathcal{A}}\rangle_\psi\UnitOp \big)\psi\rangle = \| \big(\hat{\mathcal{A}}-
\langle\hat{\mathcal{A}}\rangle_\psi\UnitOp\big)\psi  \|^2 \,,
\end{equation}
where $\langle\hat{\mathcal{A}}\rangle_\psi := \langle \psi | \hat{\mathcal{A}}\psi\rangle$.
Thus, $\Var{\hat{\mathcal{A}},\psi} >0$
for any non-zero vector of the Hilbert space $\mathcal{K}$ if and only if $\hat{\mathcal{A}}|\psi\rangle \neq
\langle\hat{\mathcal{A}}\rangle_\psi\UnitOp \psi\rangle $. The latter means that $|\psi\rangle$ is not
an eigenstate of $\hat{\mathcal{A}}$, which is the case for the wave packet \eqref{soph5} for any $\epsilon $,
and in particular for $\epsilon = 0$ due to \eqref{diver}. Thus, the singularity in this case does occur,
but it is smeared. Consequently,
the expectation values of the operators corresponding to the invariants \eqref{inv-1}--\eqref{inv-3} are singular,
but these singularities occur with non-zero fluctuations.

%%%%%%%%%%%%%%%%%%%%%%%%%%%%%%%%%%%%%%%%%%
\section{Quantum Schwarzschild black hole}
%%%%%%%%%%%%%%%%%%%%%%%%%%%%%%%%%%%%%%%%%%

Recently, we have quantized the Schwarzschild spacetime with the negative mass
parameter, $M <0$, to address the case with a naked singularity
\cite{Ola,Piotr}.  Here we consider the case with a covered singularity $M>0$.
The case $M=0$ corresponds to the Minkowski spacetime.

The Schwarzschild black hole (SBH) is a spherically
symmetric, static, solution of the Einstein field equations, parameterized by
the constant $M > 0$, having the horizon at $r=2M$, and the gravitational
singularity at $r = 0$.  It is an exact spacetime geometry of a non-rotating
``point particle'' with the mass parameter $M$ devoid of an ``internal''
dynamics \cite{Karl}.

Let us identify the SBH within our formalism.  To this end, let us consider the
line element defined by \eqref{d2} and \eqref{dd3} with $E=0$. It is clear that
constant $M$ satisfies Eq. \!\eqref{dd5}. To see that it corresponds to the SBH,
we argue as follows. Evaluating the Einstein tensor with so obtained metric
leads to zero, which means that it corresponds to the vacuum case. The Birkhoff
theorem implies that this spherically symmetric vacuum solution must be
diffeomorphic to the Schwarzschild metric (for more details see \cite{LLB}).

It results from  \eqref{KKK}--\eqref{OS4} that the only non
vanishing invariant is the Kretschmann scalar $K = 48 M^2 r^{-6}$. In our recent
paper \cite{Ola} we have shown that quantization smears the singularity indicated
by the Kretschmann scalar avoiding its localization in the configuration space.
Here we can obtain quite similar result using the quantum state \eqref{soph5}
with $\epsilon = 0$. Applying the reasoning following Eq.~\eqref{diver}
leads to the result
\begin{equation}\label{Sch1}
  \mathrm{var}(\hat{K}; \psi_{\eta_1,\eta_2}) > 0,
\end{equation}
where $\hat{K}$ is the quantum operator corresponding to the Kretschmann scalar $K$.
Thus, the singularity in this case does occur, but it is smeared.

However, making use of the quantum state \eqref{soph5} with $\epsilon > 0$, we obtain the
result \eqref{eq:ExpAOp} specialized to the case of the Kretschmann operator.
Therefore, we conclude that the expectation value of the Kretschmann operator is finite,
which resolves the classical singularity problem.

%%%%%%%%%%%%%%%%%%%%%
\section{Conclusions}\label{Con}
%%%%%%%%%%%%%%%%%%%%%

We have shown in the paper that the evolution of the considered OS dust star model consists
of three phases: classical collapse towards the gravitational singularity, quantum evolution,
and classical expansion away from the singularity.
The quantum evolution consists of quantum collapse, strongly quantum regime, and quantum
expansion. The quantum collapse and expansion are described by the evolution of the
expectation value of the position operator. The strong quantum regime may present regular
quantum bounce or smeared singularity.

We have obtained the above scenario by making use of the vector state in the form of the
wave packet with compact support parameterized by  two real parameters $\eta_1$ and $ \eta_2$.
Thus, these vector states represent a wide class of quantum states in
considered Hilbert space $\StateSpace{K}'$ (see App. C). Therefore, our resolution of the classical gravitational
singularity of the OS model is not generic, but possible within considered quantum system
ascribed to that classical one.

Separate interpretation needs the issue of the horizons of considered isolated gravitational
system. As the system is  spherically symmetric, it possesses the horizon
that is formed during the first classical phase of its evolution if the dust density
is high enough  (indirectly due to the Birkhoff theorem). The second expression for the horizon,
derived from Eq.~\eqref{two-hor}, following the bounce does not have so clear meaning.
We suggest, it may describe the disappearance of the first horizon created
before the bounce. In such case, the bouncing time of considered black hole can be estimated
by $\frac{8}{3}M_0$. The quantum evolution may describe the so called black-to-white hole transition
(see, e.g., \cite{Mux} and references therein).

We present the quantum Schwarzschild spacetime for the positive mass parameter, $M>0$,
promised to be carried out in our recent paper \cite{Ola} concerning the case $M<0$.
That paper concerns  the naked singularity. Here we are concerned with the
covered singularity, i.e. the black hole case. The resolution of the classical
singularity is done in the way quite similar as for the OS spacetime owing to the fact
that the former is a special case of the latter.

Our quantum description of the collapsing dust star corresponds to the flat sector
of the dust collapse presented in \cite{Cla} (while described in comoving coordinates).
However, we apply different parametrization of the affine group. It is worth to mention
that different parameterizations may lead to unitarily inequivalent quantum theories \cite{AWT}.
In our parametrization the expectation values of elementary observables calculated
in coherent states coincide with corresponding classical observables. Therefore, we use
the quantum states in the form of wave packets. They enable obtaining the quantum bounce
of the space operator.  All the states parameterized by the parameters like
deltas'and epsilon in Eq.~\eqref{soph5} satisfy required physical condition and  improve
flexibility of the model. In principle, such parameters  can be determined either
by observational data or more realistic model of matter field.
Another difference is that we have analysed the evolution
of the curvature invariants. Obtained expectation values
of corresponding quantum operators are either singular but smeared, or regular.

In both approaches, ours and that of \cite{Cla}, the bouncing time of considered quantum
black hole is numerically similar.
In both papers, the issue of possible Hawking radiation is relegated to future papers.

In this paper we have used an idea of time to be a
quantum observable, similarly to the space position quantum observable. A
proposal of treating time on the same footing as other quantum observables
is recently published as the Projection Evolution approach (PEv) in
\cite{gozdz20}. In the present paper we do not apply the full PEv formalism which
allows, in principle,
to get detailed dynamics of the system by construction of the so called
evolution operators. Instead, we build a family of  possible, but not all,
evolution scenarios of our OS star by constructing a family of states with
required physical properties. One of the most important feature of our wave
packets is that in a limit of very narrow functions $\psi_{\eta_1,\eta_2}$ the
expectation values of $\hat{t}$,$~\hat{r}$ and $\hat{M}$ operators reproduce the
classical motion of the OS star, far from singularity. This constraint allows to
choose some possible evolution paths and calculate them as relations among
expectation values of the fundamental observables in this model, i.e.,
the time, the position and the global mass.

Our next paper will concern the quantum system ascribed to the LTB model of
collapsing star. Within this model one can consider the inclusion of realistic
features of matter field such as pressure and inhomogeneity in density distribution,
and reasonable equation of state. Considering the cases with covered singularity
and naked singularity separately, will enable making deeper insight into the issue
of the quantum bounce.

%%%%%%%%%%%%%%%%%%%%%%%%%%%%%%%%%%%%%%%%%%%%%%%%%%%%%%%%%%%%%%%%%%%%%%%%%%%%%%%%%%%%%%%%%%%%%%%%%%%%%%%%%
\acknowledgments We would like to thank Claus Kiefer, Jerzy Lewandowski and Edward Wilson-Ewing for helpful
discussions. JJO acknowledges the support of the National Science Centre (NCN, Poland) under the Sonata-15
research grant UMO-2019/35/D/ST9/00342.
%%%%%%%%%%%%%%%%%%%%%%%%%%%%%%%%%%%%%%%%%%%%%%%%%%%%%%%%%%%%%%%%%%%%%%%%%%%%%%%%%%%%%%%%%%%%%%%%%%%%%%%%%

\appendix

%%%%%%%%%%%%%%%%%%%%%%%%%%%%%%%%%%%%%%%%
\section{General analytical solution}
%%%%%%%%%%%%%%%%%%%%%%%%%%%%%%%%%%%%%%%%

The equations of \eqref{com1}--\eqref{com2} have analytical solutions
depending on the sign of $E$ function. They can be written in the parametric
form:
\begin{eqnarray}
  &E<0&,\;\;r=-\frac{M}{E}\left(1-\cos\eta\right),\;\;\eta - \sin\eta = \frac{\left(-E\right)^{3/2}}{M}\left(t-t_B(R)\right), \nonumber \\
  &E=0&,\;\;r=\left(\frac{9}{2}M\left(t-t_B(R)\right)^2\right)^{1/3}, \label{solutions} \\
  &E>0&,\;\;r=\frac{M}{E}\left(\cosh\eta-1\right),\;\;\sinh\eta-\eta = \frac{\left(E\right)^{3/2}}{M}\left(t-t_B(R)\right),\nonumber
\end{eqnarray}
where $\eta$ is an auxiliary parameter and $t_B(R)$ is the integration function. The cases $E>0$,  $E<0$ or $E=0$ correspond
to unbounded, bounded, and marginally bounded models, respectively (see, e.g. \cite{PlebanskiKrasinski} for more details).

%%%%%%%%%%%%%%%%%%%%%%%%%%%%%%%%%%%%%%%
\section{Coordinate transformations}
%%%%%%%%%%%%%%%%%%%%%%%%%%%%%%%%%%%%%%%

In order for this article to be self-contained we present the transformations
between different coordinate systems used throughout the text (see,
e.g. \cite{Gieseletal}). Let us recall that we denote the generalised
Painlev\'e-Gullstrand coordinates (GPG) as $(t,r,\theta,\phi)$, and co-moving
synchronous coordinates as $(\tau,R,\theta,\phi)$. We can derive the proper
coordinate transformation by direct inspection of the line elements. The LTB
metric in co-moving synchronous coordinates reads
\begin{equation}
\rmd s^2 = -\rmd \tau^2
+ \frac{\left(\frac{\partial r}{\partial R}\right)^2}{1+E}
\rmd R^2 + r^2 \rmd\Omega^2 \;.
\end{equation}
The GPG coordinates are defined via $t=\tau$ and  $r=r(R,\tau)$. Since
\begin{equation}
\rmd r = \frac{\partial r}{\partial R} \rmd R
+ \frac{\partial r}{\partial \tau}\rmd \tau \;,
\end{equation}
we get by direct substitution
\begin{equation}
\rmd s^2 = -\rmd \tau^2
+ \frac{\left(\rmd r - \frac{\partial r}{\partial \tau}\rmd \tau \right)^2}{1+E}+r^2\rmd \Omega^2 \;.
\end{equation}
Using \eqref{com1} (with the negative sign of the root) we arrive at
\begin{equation}
 \rmd s^2 = - \rmd t^2 +
\frac{(\sqrt{2 M/r + E}\, \rmd t + \rmd r)^2}{1+E} + r^2 \rmd \Omega^2 \, ,
\end{equation}
where we re-inserted the $t$ coordinate. It is worth noticing that the field
equations \eqref{dd3}--\eqref{dd4} and \eqref{com1}--\eqref{com2} are equivalent (multiply
both sides of \eqref{dd3} by $\left(\partial M/\partial r\right)^{-1}$.

%%%%%%%%%%%%%%%%%%%%%%%%%%%%%%%%%%%%%%%%%%%%%%%%%%%%%%%%%%%%%%%%%%%%%%%%%
\section{The state spaces} \label{app:StateSpaces}
%%%%%%%%%%%%%%%%%%%%%%%%%%%%%%%%%%%%%%%%%%%%%%%%%%%%%%%%%%%%%%%%%%%%%%%%%

Within the affine quantization method, we deal with two Hilbert spaces. The first
one $\StateSpace{H}=L^2(\RNumb_+,d\nu(x))$ is the carrier space of the unitary
representation $U(t,r)$ of the affine group $\Aff$. The second Hilbert space
$\StateSpace{K}=L^2(\Aff,d\mu(t,r))$, of square integrable functions on the
affine group $\Aff$ with the scalar product
\begin{equation}
\label{eq:HtoKScProd}
\ScProd{\Phi_2}{\Phi_1}:= \frac{1}{A_\phi} \int_{\Aff} d\mu(t,r) \, ,
\Phi_2(t,r)^\star \Phi_1(t,r)
\end{equation}
provides a physical interpretation of the vectors $\Psi \in \StateSpace{H}$
using the mapping $\iota_G: \StateSpace{H} \to \StateSpace{K}$ defined as:
\begin{equation}
\label{eq:HtoK}
\iota_G(\Psi):= \BraKet{t,r}{\Psi} \, ,
\end{equation}
where $\BraKet{t,r}{\Psi} \in \StateSpace{K}$ is a probability amplitude of
localization of the state $\Psi \in \StateSpace{H}$ in the coherent state
$\Ket{t,r}$. The coherent states, in turn, are quantum counterparts of the
classical configuration space points, where
\begin{equation}
\label{eq:PointCS}
\kappa: T \ni (t,r) \to \Ket{t,r} \in \StateSpace{H}
\end{equation}
is a one-to-one mapping between classical and quantum configuration spaces.

The mapping \eqref{eq:HtoK} is an unitary transformation of the Hilbert space
$\StateSpace{H}$ into a subspace $\StateSpace{K}'$ of the Hilbert space
$\StateSpace{K}$, i.e., it is not a one-to-one transformation between the spaces
$\StateSpace{H}$ an $\StateSpace{K}$. Using standard decomposition of any vector
$|\Psi \rangle \in \StateSpace{H}$ into the coherent states $\Ket{t,r}$, one
gets the equality of the scalar products in both spaces $\StateSpace{H}$ and
$\StateSpace{K}$
\begin{equation}
\label{eq:HtoKUnitary}
\BraKet{\Psi_2}{\Psi_1} = \ScProd{\iota_G(\Psi_2)}{\iota_G(\Psi_1)}
\end{equation}
for all $\Psi_2,\Psi_1 \in \StateSpace{H}$.

For further considerations let us introduce a projection operator
$\hat{\mathcal{N}}$ in the state space $\StateSpace{K}$ defined as
\begin{equation}
\label{eq:OverlapOp}
\hat{\mathcal{N}} \Phi(t,r):= \frac{1}{A_\Phi} \int_{\Aff} d\mu(t',r')
\BraKet{t,r}{t',r'} \Phi(t',r') \, .
\end{equation}
A direct calculation shows that for every $\Psi \in \StateSpace{H}$
the amplitude $\BraKet{t,r}{\Psi}$ belongs to the subspace
$\hat{\mathcal{N}} \StateSpace{K}$, i.e.,
\begin{equation}
\label{eq:NBraKetPsi}
\hat{\mathcal{N}}\, \BraKet{t,r}{\Psi} = \BraKet{t,r}{\Psi} \, .
\end{equation}
On the other hand, every $\Psi \in \StateSpace{H}$ can be expanded into the
coherent states
\begin{equation}
\label{eq:ExpansionCS}
\Psi(x)= \frac{1}{A_\phi} \int_{\Aff} d\mu(t,r)
\BraKet{x}{t,r} \BraKet{t,r}{\Psi}
\end{equation}
what implies that Eq.~\eqref{eq:ExpansionCS} defines an inverse transformation
to the unitary mapping \eqref{eq:HtoK} between the spaces $\StateSpace{H}$ and
$\StateSpace{K}'$. The subspace $\StateSpace{K}' \subset \StateSpace{K}$ is
spanned by the functions $\BraKet{t,r}{\Psi}$, which symbolically can be written
as
\begin{equation}
\label{eq:KprimSpace}
\StateSpace{K}'= \mathrm{Hilbert}\{\iota_G(\Psi): \Psi \in \StateSpace{H} \}
\, .
\end{equation}
Summarizing, we have the following hierarchy of Hilbert spaces:
$\StateSpace{K}' \subset \hat{\mathcal{N}}\StateSpace{K} \subset \StateSpace{K}$
and the spaces $\StateSpace{H}$ and $\StateSpace{K}'$ are unitarily isomorphic.

These properties allow to use either the carrier Hilbert space $\StateSpace{H}$
or the Hilbert space $\StateSpace{K}'$ to perform
calculations, depending on context, while keeping their appropriate physical
interpretations.

In the case of the spherical symmetry, $(t,r) \in T$ denotes the time $t$  and the two
dimensional sphere with the radius $r$.  That point is represented by the
coherent state $\Ket{t,r} \in \mathcal{H}$ for which the expectation values of the
time and the radius operators are $\Bra{t,r}\hat{t}\Ket{t,r}=t$ and
$\Bra{t,r}\hat{r}\Ket{t,r}=r$, respectively. The scalar product
$\BraKet{t,r}{\Psi}$, where $\Ket{\Psi} \in \StateSpace{H}$, is the  probability
amplitude of the localization of the system at the time $t$ in a form of a three
dimensional spherical object with the  radius  $r$.

%%%%%%%%%%%%%%%%%%%%%%%%%%%%%%%%%%%%%%%%%%%%%%%%%%%%%%%%%%%%%%%%%%%%%%%%%

\end{document}